\documentclass[a4paper]{jpconf}
\usepackage{graphicx}
\usepackage{iwsmihead}
\usepackage{ogonek}
\usepackage{wrapfig,wrapft}

\def\(({\left(}
\def\)){\right)}                       
\def\[[{\left[}
\def\]]{\right]}

\newcommand{\be}{\begin{equation}}
\newcommand{\ee}{\end{equation}}
\newcommand{\bea}{\begin{eqnarray}}
\newcommand{\eea}{\end{eqnarray}}

\begin{document}

\title{Phase Transitions and Computational Difficulty in 
Random Constraint Satisfaction Problems}

\author{Florent Krz\k{a}ka{\l}a$^1$ and Lenka Zdeborov\'a$^2$}

\address{ $^1$ PCT, UMR Gulliver 7083 CNRS-ESPCI, 10 rue Vauquelin, 75231 Paris, France \\
 $^2$ LPTMS, UMR 8626 CNRS et Univ. Paris-Sud, 91405 Orsay CEDEX, France\\}

\ead{fk@espci.fr}

\begin{abstract}
  We review the understanding of the random constraint satisfaction problems, focusing on the $q$-coloring of large random
  graphs, that has been achieved using the cavity method. We also discuss the
  properties of the phase diagram in temperature, the connections with the
  glass transition phenomenology in physics, and the related algorithmic
  issues.
\end{abstract}

\section{Introduction}
\begin{wrapfigure}{r}{5.7cm}
  \includegraphics[width=5.5cm,height=4.5cm]{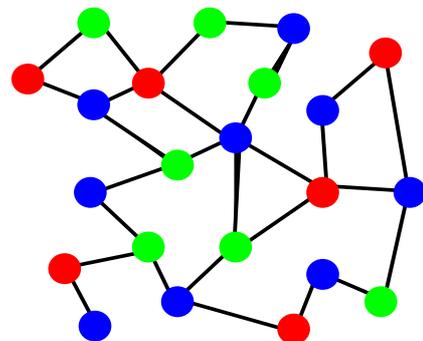}
  \caption{Example: a proper 3-coloring of a small graph.}
  \label{example}
\end{wrapfigure}

Spin glass theory has a large and probably initially unexpected impact
on some problems far from condensed matter physics and one example of
such spectacular outcome is the application of statistical physics
ideas to combinatorial optimization~\cite{GaJo} and of the concept of
phase transitions to the probabilistic analysis of Constraint
Satisfaction Problems (CSPs)~\cite{MPV,MonassonZecchina99,MPZ}.  Given
a set of $N$ discrete variables subject to a set of $M$ constraints, a
CSP consists in deciding if there exists an assignment of the
variables satisfying all the constraints.  This is a generic setting
that is currently used to tackle problems as diverse as, among others,
error-correcting codes, register allocation in compilers or genetic
regulatory networks. The class of NP-complete problems~\cite{GaJo},
for which no algorithm is known that guarantees to decide
satisfiability in a time polynomial in $N$, is particularly
interesting.  Well-studied examples of such problems are the
satisfiability of boolean formulas (SAT), and the $q$-coloring problem
($q$-COL, see figure \ref{example}) that we shall discuss here. Given a
graph with $N$ vertices and $M$ edges connecting certain pairs of
them, and given $q$ colors, can we color the vertices so that no two
connected vertices have the same color?

Crucial empirical observations were made when considering the ensemble of
random graphs with a given average vertex connectivity $c$: while below a
critical value $c_s$ a proper $q$-coloring of the graph exists
with a probability going to one in the large size limit, it was found that
beyond $c_s$ no proper $q$-coloring exists asymptotically. This sharp threshold
(which appears in other CSPs such as K-SAT and whose existence is partially
proved in~\cite{Friedgut}) is an example of a phase transition arising in
random CSPs. It was also observed empirically~\cite{CKT91,Selman} that
deciding colorability becomes on average much harder near to the coloring
threshold $c_s$ than far away from it. It is therefore natural to ask ourselves:
Can the value of the colorable/uncolorable (COL/UNCOL) phase transition be computed? Can the number of
all possible colorings be also computed?  Are there other interesting phase
transitions? Can these transitions explain the fact that solutions are
sometimes very hard to find?  Can this knowledge help us in designing new
algorithms? These questions, and their answers, are at the roots of the
interest of the statistical physics community in optimization
problems~\cite{MonassonZecchina99,MPZ}.

\section{A Potts anti-ferromagnet on random graphs}
It is immediate to realize that the $q$-coloring problem is equivalent to the
question of determining if the ground-state energy of a Potts anti-ferromagnet
on a random graph is zero or not~\cite{Kanter}. Consider indeed a graph $G =
({\cal V,E})$ defined by its vertices ${\cal V}=\{1,\dots,N\}$ and edges
$(i,j)\in {\cal E}$ which connect pairs of vertices $i,j\in {\cal V}$; and the
Hamiltonian
\be
{\cal H}(\{s\}) = \sum_{(i,j) \in {\cal E}} \delta(s_i,s_j)\, .
\label{Ham}
\ee 
With this choice there is no energy contribution for neighbors with different
colors, but a positive contribution otherwise. The ground state energy (the
energy at zero temperature) is thus zero {\it if and only if} the graph is
$q$-colorable. This transforms the coloring problem into a well-defined
statistical physics model. Usually, two types of random graphs are considered:
in the $c-$regular ensemble all points are connected to exactly $c$ neighbors,
while in the Erd\H{o}s-R\'enyi case the connectivity has a Poisson
distribution.

\section{Cavity method: Warnings, Beliefs and Surveys}
\label{cavity}
Over the last few years, a number of studies have investigated CSPs
following the adaptation of the so-called cavity method~\cite{MPV} to
random graphs~\cite{MPZ,Cavity}. It is a powerful heuristic tool
---whose exactness is widely accepted but has still to be rigorously
demonstrated--- equivalent to the replica method of disordered
systems~\cite{MPV}. Its main idea lies in the fact that a large random
graph is locally tree-like, and that an iterative procedure known in
physics as the Bethe-Peirls method can solve exactly any model on a
tree (such models are often qualified as ``mean field'' in physics).
Interestingly, it was realized~\cite{Yedidia} that an equivalent
formalism has been developed independently in computer
science~\cite{BP-Pearl}, where it is called Belief Propagation (BP,
which conveniently enough, may also stands for Bethe-Peirls). Defining
$\psi^{i\to j}_{c} (c=1,..,q)$ as the probability that the spin $i$
has color $c$ in absence of the spin $j$ (the ``belief'' that the spin
$j$ has on the properties of the spin $i$), BP reads
\begin{eqnarray}
  \psi_{c}^{i\to j} = \frac{1}{Z_0^{i\to j}} \prod_{k\in N(i)\setminus\{j\}} \left(1- \psi_{c}^{k\to i} \right) \label{update_zero}
\end{eqnarray}
where $Z_0^{i\to j}$ is a normalization constant and the notation
$k\in N(i)\setminus\{j\}$ means the set of neighbors of $i$ except
$j$. From a fixed point of these equations, the complete beliefs in
presence of all spins can be also computed.  They give, for each
vertex, the probability of each color from which other quantities, as
for instance the number of solutions, can be computed. A simpler
formalism, called Warning Propagation (WP), restricts itself to frozen
variables ({\it i.e.~}to variables for which only one color can
satisfy the constraints). However, WP does not allow to compute the
number of solutions, only their existence, but is definitely simpler
to handle.

It was soon realized, however, that these methods developed for trees could
not be used straightforwardly on all random graphs because of a non-trivial
phenomenon called clustering~\cite{Cavity,BMW} (for which rigorous results are
now available, see \cite{Thierry}). Indeed, while for graphs with very low
connectivities all solutions are ``connected'' ---in the sense that it is easy
with a local dynamics to move from one solution to another--- they 
regroup into a large number of disconnected clusters for larger connectivities. 
It can be argued that each of these clusters corresponds to a
different fixed point of the BP equations, so that a survey over
the whole set of the fixed points should be performed. This can be done in the
cavity method by the now famous Survey Propagation (SP) equations~\cite{MPZ}
which, in the physics language, correspond to the Parisi's one-step Replica
Symmetry Breaking (RSB) scheme~\cite{MPV}.  Within this formalism, the number
of clusters (which behaves as ${\cal{N}}=e^{N\Sigma}$, where $\Sigma$ is a fundamental quantity called the {\it complexity})
and their sizes (the number of proper colorings inside the cluster) can be
determined.

This formalism has been applied on the SAT~\cite{MPZ} and
COL~\cite{Coloring,ColoringFlo} problems in the limit of infinitely large
graphs. These cutting edge studies were however restricted to SP applied to
the clusters corresponding to fixed points of WP and not to those of BP.
Although this already allowed the correct computation of the COL/UNCOL
transition and the development of a powerful algorithm \cite{MPZ}, it meant
that the description of the clustered phase was only partial and this resulted
in a number of problems and inconsistencies that stayed unanswered until very
recently. These issues have been today
clarified~\cite{PAL,Reconstruction,PNAS,JAMMING,COL,HARD} and we shall now
discuss this new understanding.

\section{The phase diagram of the coloring problem on a random graph}
\label{phase}
\begin{figure}
\begin{center}
\includegraphics[width=\linewidth]{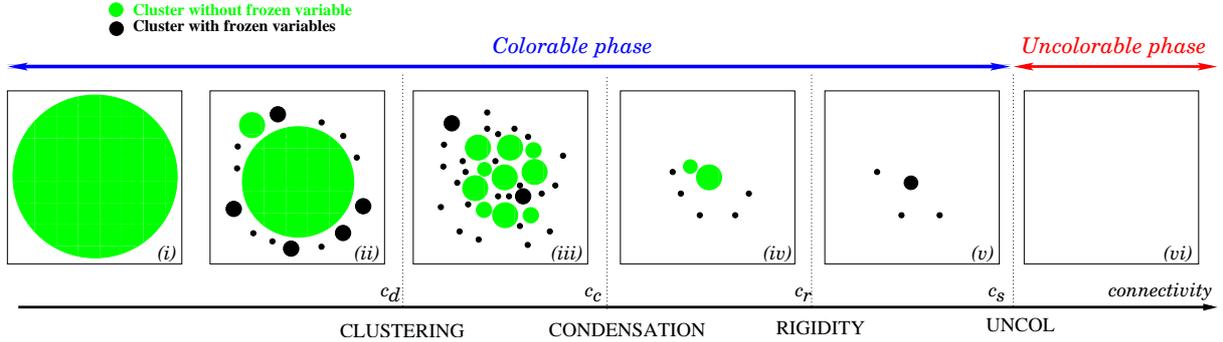}
\end{center}
\caption{\label{fig1} Sketch of the space of solutions ---colored
  points in this representation--- in the $q$-coloring problem on random
  graphs when the connectivity $c$ is increased.  (i) At low $c$, all
  solutions belong to a single cluster. (ii) For larger $c$, other clusters of
  solutions appear but a giant cluster still contains almost all solutions.  (iii)
  At the clustering transition $c_d$, it splits into an exponentially large
  number of clusters.  (iv) At the condensation transition $c_c$, most
  colorings are found in the few largest of them. (v) The rigidity transition
  $c_r$ ($c_r<c_c$ and $c_r>c_c$ are both possible depending on $q$) arises
  when typical solutions belong to clusters with frozen variables (that are
  allowed only one color in the cluster). (vi) No proper coloring exists beyond the
  COL/UNCOL threshold $c_s$.}
\end{figure}

Consider that we have $q\ge 4$ colors (the $q=3$ case being a bit particular
\cite{PAL,PNAS,COL}, as we shall see) and a large random graph whose
connectivity $c$ we shall increase. Different phases are encountered that we
will now describe (and enumerate) in order of appearance (the corresponding
phase diagram is depicted in figure \ref{fig1}).

\begin{itemize}
\item[(i)]{\bf A unique cluster exists}: For low enough connectivities, all the
  proper colorings are found in a single cluster, where it is easy to ``move''
  from one solution to another. Only one possible ---and trivial--- fixed
  point of the BP equations exists at this stage (as can be proved rigorously
  in some cases \cite{Unicity}). The entropy can
  be computed and reads in the large graph size $N$ limit
\be 
s = \frac{\log{\cal{N}}_{\rm sol}} N = \log{q} + \frac {c}{2}
\log{\((1-\frac{1}{q}\))}\, .\label{S_RS} 
\ee

\item[(ii)] {\bf Some (irrelevant) clusters appear}: As the connectivity is
  slightly increased, the phase space of solutions decomposes into an large
  (exponential) number of different clusters. It is tempting to identify that
  as the clustering transition, but it happens that all (but one) of these
  clusters contain relatively very few solutions ---as compare to whole set---
  and that almost all proper colorings still belong to one single giant
  cluster.  Clearly, this is not a proper clustering phenomenon and in fact,
  for all practical purpose, there is still only one single cluster.
  equation (\ref{S_RS}) still gives the correct entropy at this stage.

\item[(iii)] {\bf The clustered phase}: For larger connectivities, the large
  single cluster also decomposes into an exponential number of smaller ones:
  this now defines the genuine clustering threshold $c_d$\footnote{It is
    important to point out that the location of the clustering transitions was
    therefore not computed correctly when the dependence on the size of
    clusters was not taken into account. Also, different results were obtained
    previously depending on whether or not unfrozen fixed points were
    explicitly considered.}. Beyond this threshold, a local algorithm that
  tries to move in the space of solutions will remain prisoner of a cluster of
  solutions \cite{Guilhem}. Interestingly, it can be shown that the total
  number of solutions is still given by equation (\ref{S_RS}) in this phase.  This
  is because, as is well known in the replica method, the free energy has no
  singularity at the dynamical transition (which is therefore not a true
  transition in the sense of Ehrenfest, but rather a dynamical or geometrical transition in
  the space of solutions).

\item[(iv)] {\bf The condensed phase}: As the connectivity is further
  increased, a new sharp phase transition arises at the condensation threshold
  $c_c$ where most of the solutions are found in a finite number of the
  largest clusters. From this point, equation (\ref{S_RS}) is not valid anymore and becomes just an upper bound.  The entropy is non-analytic at $c_c$ therefore
  this is a genuine static phase transition. 

\item[(v)] {\bf The rigid phase}: As mentioned in section \ref{cavity}, two
  different types of clusters exist: In the first type, that we shall call the
  {\it unfrozen} ones, all spins can take at least two different colors.  In
  the second type, however, a finite fraction of spins is allowed only one
  color within the cluster and are thus ``frozen'' into this color. These {\it
    frozen} clusters actually correspond to non-trivial fixed points of BP
  {\it and} WP, while the first kind are non-trivial fixed points of BP {\it
    only}. It follows that a transition exists, that we call {\it rigidity},
  when frozen variables appear inside the dominant clusters (those that
  contains most colorings).  If one takes a proper coloring at
  random beyond $c_r$, it will belong to a cluster where a finite fraction of
  variables is frozen into the same color.  Depending on the value of $q$,
  this transition may arise before or after the condensation transition (see
  table \ref{results}).

\item[(vi)] {\bf The UNCOL phase}: Eventually, the connectivity $c_s$ is
  reached beyond which no more solutions exist. The ground state energy
  (sketched in figure \ref{fig2}) is zero for $c<c_{s}$ and then grows
  continuously for $c>c_{s}$. The values $c_s$ computed within the cavity
  formalism are in perfect agreement with the rigorous bounds \cite{Rigorous}
  derived using probabilistic methods and are widely believed to be exact
  (although they remains to be rigorously proven, but see \cite{Silvio} for a
  proof that they are at least rigorous upper bounds).
\end{itemize}
We report the values of the threshold connectivities corresponding to all
these transitions in table \ref{results} for the regular and the Poissonian
({\it i.e.~}Erd\H{o}s-R\'enyi) random graphs ensembles. Notice that the
$3$-coloring is peculiar because $c_d=c_c$ so that the clustered phase is
always condensed in this case. In view of this rich phase diagram, it is
important to get an intuition on the meaning and the properties of these
different phases and, in this respect, it is interesting before entering the
algorithmic implications to discuss the analogies with the glass transition.
\begin{center}
\begin{table}
\begin{minipage}{0.45\linewidth}
\begin{tabular}{|c||l|l|l|l|}
\hline
q & $c_d$ & $c_r$ & $c_c$ & $c_s$ \\
\hline
\hline
3 & $5^+$ & - & 6 & 6 \\
\hline
4 & 9 & - & 10 & 10 \\
\hline
5 & 14  & 14 & 14 & 15 \\   
\hline
6 & 18 & 19 & 19 & 20  \\   
\hline
7 & 23 & - & 25 & 25  \\   
\hline
8 & 29 & 30 & 31 & 31  \\   
\hline
9 & 34 & 36 & 37 & 37  \\   
\hline
10 & 39 & 42 & 43 & 44  \\   
\hline
\end{tabular}
\end{minipage}
\begin{minipage}{0.5\linewidth}
\begin{tabular}{|c||l|l|l|l|}\hline
  q & $c_d$ & $c_r$ &  $c_c$ & $c_s$ \\
  \hline \hline
  3 & 4 & 4.66(1) & 4 & 4.687(2) \\
  \hline
  4 & 8.353(3) & 8.83(2) & 8.46(1) & 8.901(2) \\
  \hline
  5 & 12.837(3) & 13.55(2) & 13.23(1) & 13.669(2)  \\
  \hline
  6 & 17.645(5) & 18.68(2) & 18.44(1) & 18.880(2) \\
  \hline
  7 & 22.705(5) & 24.16(2) & 24.01(1) & 24.455(5) \\
  \hline
  8 & 27.95(5) & 29.93(3) & 29.90(1) & 30.335(5) \\
  \hline
  9 & 33.45(5) & 35.658 & 36.08(5) & 36.490(5) \\
  \hline
  10 & 39.0(1) & 41.508 & 42.50(5) & 42.93(1) \\
  \hline
\end{tabular}
\end{minipage}
\caption{\label{results}  Threshold connectivities $c_d$
  (dynamical/clustering) \cite{COL,Reconstruction,PNAS,Saad}, $c_r$
  (rigidity/freezing) \cite{COL,HARD}, $c_c$ (condensation/Kauzmann)
  \cite{COL,PNAS} and $c_s$ (COL/UNCOL) \cite{Coloring,ColoringFlo}
  for regular (left) and Erd\H{o}s-R\'enyi (right)
  random graphs. In
  the large $q$-limit, one finds in both cases that \cite{COL,PNAS}: $c_r = q
  [\log q + \log\log q + 1 + o(1)]$, $c_c = 2q \log{q} -  \log{q} - 2 \log{2} +
  o(1)$ and \cite{ColoringFlo}: $c_s=  2q \log{q}-\log{q} -1 + o(1)$.}
\end{table}
\end{center}

\section{A detour into the ideal glass transition phenomenology}
\label{glass_t}
To those familiar with the replica theory and the mean field theory of
glasses, the phenomenology depicted in the former section should look
familiar: these successive transitions are indeed very well known in
the picture of the ideal glass transition~\cite{GLASS}.  This striking
analogy is in fact quite natural since, despite the fact that there is
no disorder in the interactions in Hamiltonian (\ref{Ham}), the
frustration due to the loops in the random graph makes the model
behaving like a disordered ``anti-ferromagnetic'' Potts spin
glass~\cite{FiniteT} and such models are known to display the glassy
phenomenology~\cite{Kanter,GLASS}.

\begin{figure}
\includegraphics[width=25pc]{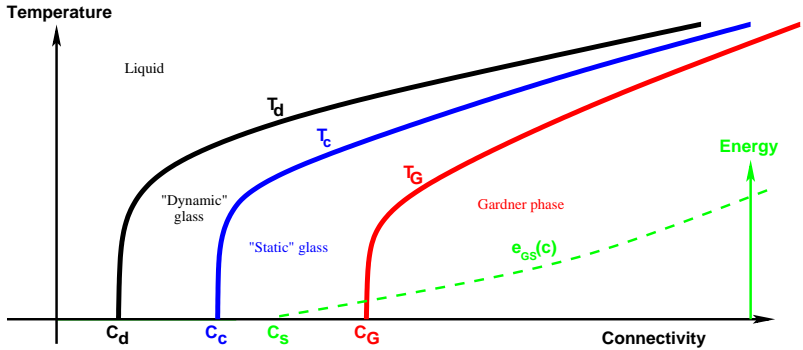}\hspace{1pc}%
\begin{minipage}[b]{12pc}\caption{\label{fig2} Sketch of the phase
    diagram in the coloring problem at finite temperature (from
    \cite{FiniteT}). At $T_d$, the system falls out of equilibrium
    (``dynamic'' transition). At $T_c$ the system undergoes a
    ``static'' glass transition. Finally, at $T_G$, a Gardner
    transition appears. $e_{GS}$ represents the ground state energy.}
\end{minipage}
\end{figure}

The phase diagram obtained on Poissonian random graphs with average
connectivity $c$ for $q\ge4$ is sketched in figure \ref{fig2} (the
$q=3$ model is slightly different as, again, $T_d=T_c$). At high
temperature the system behaves as a liquid (or a paramagnet in the
language of magnetic systems).  Below a temperature $T_d$ a first
transition ---called ``dynamical''--- happens and the system falls out
of equilibrium. For $T<T_d$ it is not possible for a physical dynamics
to equilibrate the system and the ergodicity is broken: this is due to
the appearance of exponentially many different states. However, the
would-be equilibrium properties of the problem remain similar (and in
particular, the free-energy has no singularity at this
temperature). Only at temperature $T_c$ the free energy is non
analytic and a true ``static'' glass transition happens, called the
Kauzmann transition~\cite{GLASS}.  In this phase only a finite number
of states does matter at a given connectivity.  Finally, for larger
connectivities, a third phenomenon is observed as the temperature is
further lowered, called the Gardner transition \cite{Gardner,Luca}.
It is a transition towards a more complicated phase, similar to the
one found in the celebrated solution of the Sherrington-Kirkpatrick
model \cite{MPV}. The fact that the Gardner transition arises for
connectivities {\it larger} the COL/UNCOL one is very important in
this respect: it shows that the study of this phase, that requires a
more involved cavity formalism (and probably further RSB), does not
seem to be needed in the colorable phase\footnote{The expert reader
  might find this puzzling, as many papers stated that the simple
  ``one replica symmetry breaking'' \cite{MPZ,Cavity} solution was
  unstable towards a more complex solution in some region of the
  COL/SAT phase \cite{SATSTAB,ColoringFlo}.  However, these results
  were obtained neglecting the role of the sizes of the clusters:
  while in some cases {\it most} clusters are indeed unstable, our
  studies \cite{FiniteT} indicate that the {\it relevant ones} seem
  always stable in the COL phase (although the cases of $3-$COL and
  $3-$SAT might be problematic, see \cite{COL,FiniteT}).}. We also now
recognize that the ``clustering'' and the ``condensation'' transitions
in the coloring problem are just the zero temperature relics of the
dynamic and Kauzmann transitions at finite temperatures.

A similar connection with the physics of glassy system
can also be obtained directly at zero temperature via the jamming
phenomenology~\cite{JAMMING} where the density of constraints (in this case
the volume of some non-overlapping spheres in a box of fixed volume) are
increased and where a dynamical transition is first met while some authorized
configurations exist much beyond this point (see for instance \cite{Liu} and
references therein).
We thus see that the coloring problem on random graphs translates into a very
general mean field model of a complex liquid. This convergence of interest
between different disciplines is quite interesting in itself and allows to
discuss a number of matters, as we shall now see.

\section{Onset of hardness for local search algorithms}
\label{algo}
The properties of the phase diagram we just discussed are based on analytical
computations through the cavity method. We would like to discuss now what are
the implications of these different phases on the performance of simple local search 
algorithms that try to find a solution. This is, however, a much harder
subject to handle analytically and we shall thus leave the field of analytical
computations to enter the one of phenomenology. Still, the behavior of an
algorithm trying to find a solution is reminiscent of the behavior of the
physical dynamics in glassy systems, and we can at least exploit this analogy
in order to get an intuition for the problem that we can later confirm with
numerical simulations.

It is first tempting to identify the point $c_d$, where a physical Monte-Carlo
dynamics gets trapped into a cluster, with the onset of computational
hardness\footnote{When the clustering phenomena was discovered in CSPs, it was
  indeed initially conjectured to be responsible for the onset of hardness for
  local search strategies~\cite{BMW,MPZ} and the clustered phase was named the
  ``hard phase''. However, some local algorithms were found to easily beat
  the threshold (see \cite{John} for SAT and \cite{COL} for COL).}.  However, a
second moment though indicates that this should not be the case: In the glass
transition phenomenology, it is well known that, although the system falls out
of equilibrium beyond $T_d$, its energy can be further lowered by lowering the
temperature, or just by waiting a bit longer \cite{JAMMING}.  In short: the
fact that dynamics is prisoner of a given region of the set of possible
configurations does not mean that no solution can be found in this region.
Although $c_d$ is indeed a sharp transition for the Monte-Carlo sampling, there is no reason, a priori, to experience difficulties if one just
want to find one solution beyond this point. This is particularly transparent
in the analysis and the algorithm introduced in \cite{JAMMING} (and directly
inspired from the analogy with jamming \cite{Liu}):
\begin{enumerate}
\item Start with a graph of connectivity $c$ and a proper coloring. 
\item Increase the density of constraint by adding a link in the graph.
\item Use a simple algorithm in order to solve the contradiction introduced by
  the link. When it is done, go back to step (i).
\end{enumerate}
By applying this strategy, starting from scratch ({\it i.e.~}from a graph with
$N$ vertices and {\it no} link), the set of all proper colorings undergoes the
successive transitions described in figure \ref{fig1} as connectivity increases.
When the dynamical transition is reached, one is trapped inside a cluster of
solutions, but this is not really a problem as one is still free to move
inside the cluster. As more links are added, the cluster size is
 decreasing continuously but while it still exists, the local algorithm should
be in principle able to find solutions nearby. Only for larger connectivities, 
when the cluster gets frozen, it disappears and consequently the algorithm stops. It was shown
in Ref.~\cite{JAMMING} through numerical simulations that this strategy,
using the Walk-COL~\cite{COL} algorithm for step (iii), is indeed
efficient, and linear in $N$, much beyond the dynamical transition.

The reason why this recursive strategy becomes inefficient when the cluster in which
the dynamics is trapped freezes is the following: if a link is put between
two vertices frozen in the same color, it is impossible to satisfy
the constraints while remaining in the cluster. As opposed to the unfrozen
clusters, the frozen clusters thus have a finite probability to disappear when a
new link is added. A cavity-like analysis \cite{HARD}, confirmed by numerical
data \cite{JAMMING}, actually shows that the number of changes that the
algorithm must perform, in order to solve the contradictions imposed by the
addition of new links, increases with the connectivity and diverges when the
frozen variables appear. The source of difficulties is therefore not the
clustering phenomenon in itself, but rather the appearance of frozen
variables. This makes the analysis and the prediction of a EASY/HARD threshold
much harder since (as one can see on figure \ref{fig1}) clusters of different sizes freeze at different connectivities, although a
connectivity $c_*\ge c_r$ exists where all clusters are frozen, thus putting a
strict bound to the efficiency of this procedure.
\begin{figure}
\begin{minipage}{18pc}
\includegraphics[width=18pc,height=5cm]{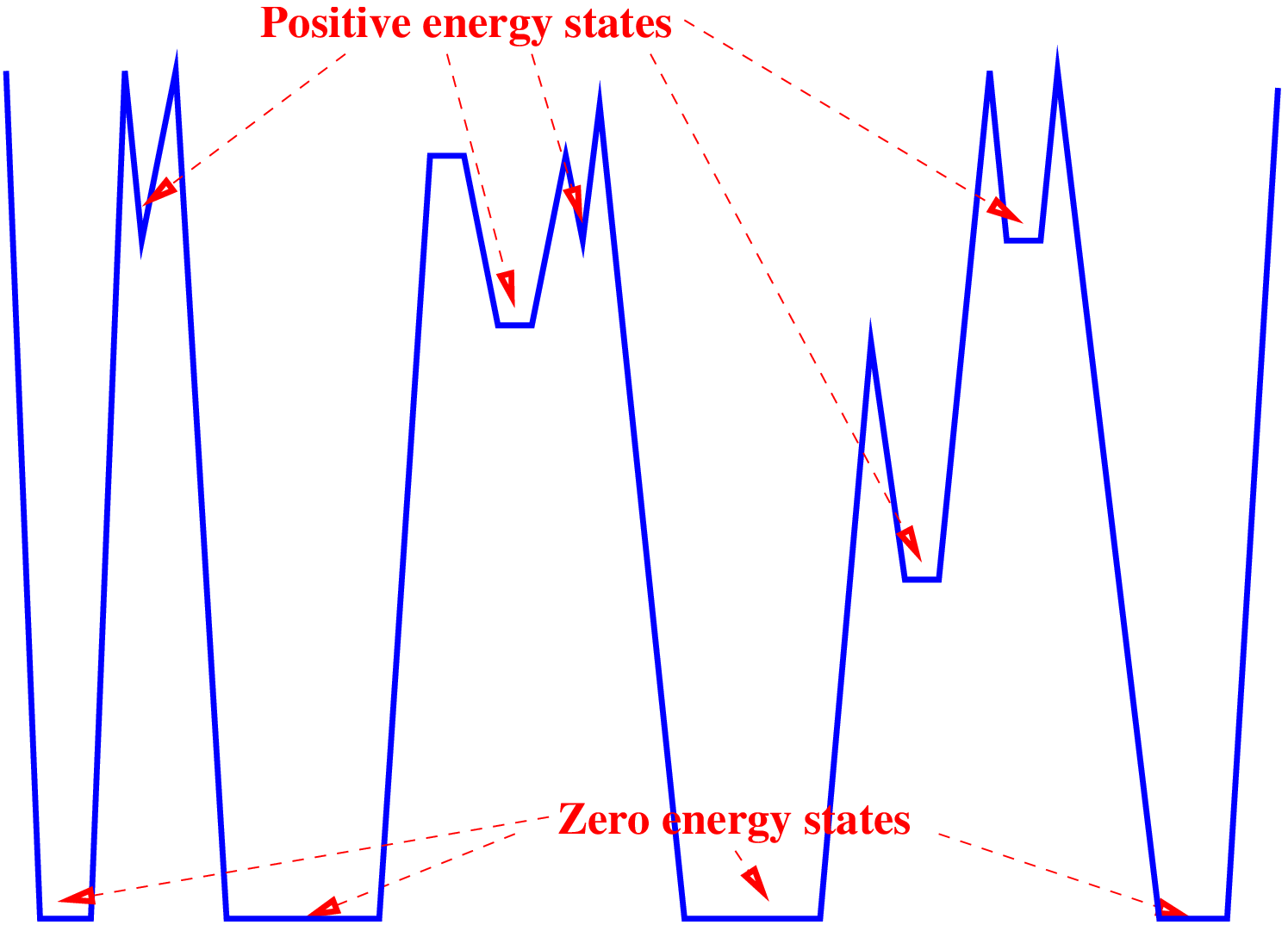}
\caption{\label{land1} Artist's view of the energy landscape for low
  connectivities $c>c_d$: a region dominated by canyons that reach the
  ground-states.}
\end{minipage}\hspace{2pc}%
\begin{minipage}{18pc}
\includegraphics[width=18pc,height=5cm]{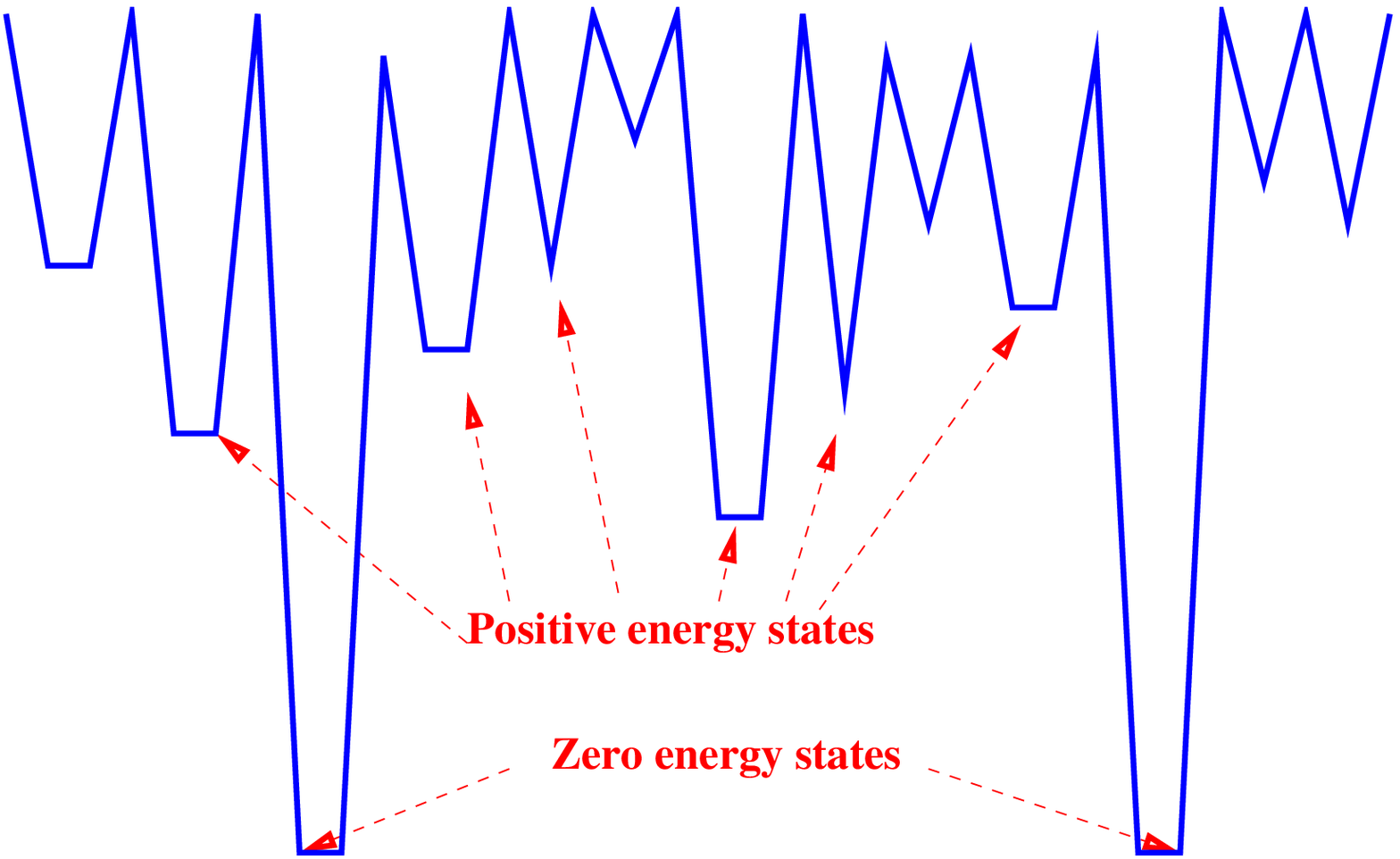}
\caption{\label{land2} Artist's view of the energy landscape for large
  connectivities $c>c_d$: a region dominated by high mountains and deep
  valleys.}
\end{minipage} 
\end{figure}

Interestingly, even non-incremental algorithms may also pass the $c_d$
threshold \cite{John,COL}, and that might come as a surprise for those
having in mind the ``rugged'' many-valley energy landscape picture of
spin glasses. This apparent paradox can be clarified by the following
considerations: It is possible that at lower connectivity $c>c_d$ the
energy landscape is dominated by deep canyons (figure \ref{land1}),
where it is in principle easy to go down as one has just to jump
ahead!  At larger connectivities a more rugged region with many deep
valleys and high mountains is found (figure \ref{land2}) in which case,
as any mountain-hiker will undoubtly know, it takes some time to go to
the deepest valley because many hills have to be climbed first. This
difference in behavior might explain the ``unreasonable efficiency''
of local algorithm \cite{Selman} and the performance of the annealing
procedure beyond $c_d$ \cite{Saad}.

\begin{figure}
\begin{minipage}{25pc}
\includegraphics[angle=270,width=25pc]{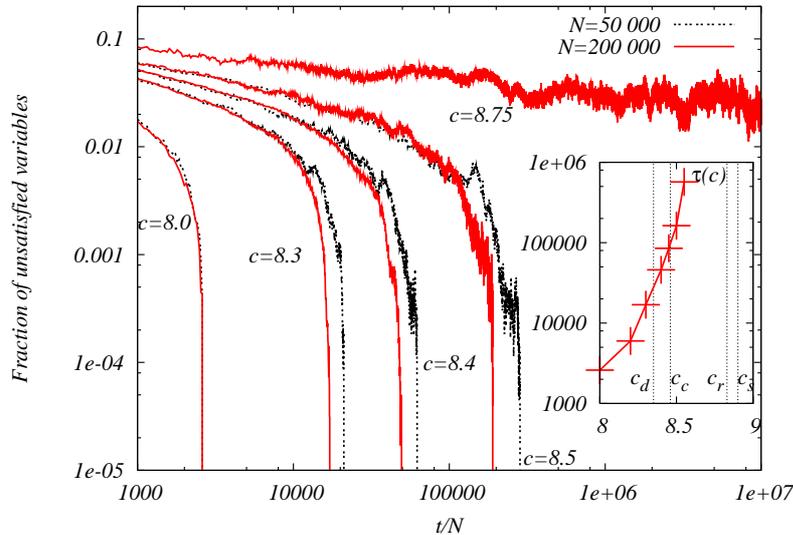}\hspace{2pc}%
\end{minipage}
\begin{minipage}[]{13pc}\caption{\label{WALK} Fraction of unsatisfied
    variables versus the number of attempted flips of the Walk-COL algorithm
    divided by the size of the graph in the $q=4$ coloring: 
    Walk-COL \cite{COL} algorithm is able to find some solutions 
    in the clustered phase for low $c$ (in the canyon-like
    region), but get trapped in the high energy 
    valleys for larger $c$. 
    Inset: estimated time $\tau=t/N$ needed to find a solution versus connectivity.}
\end{minipage}
\end{figure}

To further illustrate this
point, consider the Walk-COL algorithm introduced in \cite{COL} (and adapted
from a similar one in SAT \cite{John}) defined by the following procedure
\begin{enumerate}
\item[(i)] Randomly choose a spin that has the same color as at least one of its neighbors.
\item[(ii)] Change randomly its color. Accept this change with probability
  one if the number of unsatisfied spins has been lowered, otherwise accept
  it with probability $p$ (this is a parameter that has to be tuned for
  better efficiency).
\item[(iii)] If there are unsatisfied vertices, go to step (i) unless the
  maximum running time is reached.
\end{enumerate}
This algorithm can easily find colorings for large sizes in linear time beyond
$c_d$ \cite{COL}, but certainly not too close to the UNCOL transition where it
gets trapped at higher energies (see figure \ref{WALK}).

So far, there are few analytical results about the energy landscape in this
problem and it is likely that this will be the subject of further studies.  It
is unfortunately very hard to say for which connectivities the landscape goes
from canyons-dominated to mountains-dominated as this may not be a sharp
transition and more a matter of ---certainly algorithm-dependent--- basins of
attraction. The rigidity transition for typical clusters is certainly a good
candidate as a crossover in behavior (as is the connectivity where {\it
  all} clusters are frozen).

To conclude, one sees that although the algorithmic issues are indeed
more difficult to handle than the phase diagram, at least two important
points can already be made: First, the dynamical transition does not
correspond to the onset of hardness, and second, the source of difficulty
seems more to be related with the appearance of frozen variables.

\section{Message Passing and Decimation}
The class of local search algorithms is only one part of the story. A
different class, where messages are exchanged through the
nodes of the graph, was proven to be very efficient in computer science.  BP
and WP are examples of such procedures and it is thus interesting to use the
information given by the cavity analysis to discuss their performance in
estimating the marginals ---or other informations--- in the problem. A major
outcome of the last years has also been the application of SP as a
message passing (MP) \cite{MPZ}.

From the information given by the fixed point of the MP, an algorithm can be
defined in the following way \cite{MPZ}: (i) run the MP to obtain some
information, and (ii) fix (decimate) some variables according to this information. This sequence is
iterated until a solution is found, or until a contradiction is met. The two
parts are quite independent and each of them can be changed separately (for
instance a self consistent re-enforcement has been tried instead of the
decimation with very good results \cite{FIELDS}).

According to the cavity formalism, the iterative fixed point of BP correctly
estimates the marginals until the condensation at $c_c$ \cite{PNAS} 
(which, conveniently,
is very close to $c_s$). Indeed the application of BP plus decimation is
numerically efficient in finding solutions for both SAT and
COL~\cite{PNAS,COL,SAT} and it would be interesting to see if this method
allows to find ``typical'' solutions beyond $c_d$, thus bypassing the problems
of Monte-Carlo algorithms. It seems that the strategy is efficient even {\it
  beyond} $c_c$ in some cases \cite{COL}, although the BP recursion does not always
converge (and when it does, it does it slowly) on the decimated graph, in which
case an imprecise and approximate 
information is obtained. These issues are thus far from
being properly understood at the present time, and we hope that new works will
be done in this direction.

A more powerful strategy in the clustered phase is to use SP to compute the
probability that a given variable is frozen is a cluster \cite{MPZ}.  Fixing
the variables which are frozen in most clusters seems a good way to decimate
the graph. Using this method in random 3-SAT allows to outperform any other
known algorithm and to find solutions of huge instances rapidly even very
close to the threshold~\cite{MPZ,FIELDS}. The method has been subsequently
adapted for the $3-$coloring in \cite{Coloring}.  Interestingly, SP behaves
better than BP on the decimated graph and has no problem of convergence. Other
strategies are possible that have not yet been exploited \cite{COL}.

The best way to extract the informations given by the fixed point of these
message passing procedures is however not yet known, nor is the limit until where these
strategies are efficient. However, this line of thoughts is certainly the most
promising way in the direction of better solving and sampling algorithms.

\section{Conclusions and perspectives}
In this paper, we considered and reviewed some aspects of the $q-$coloring
of large random graphs. We discussed the properties of the set of
solutions and the different sharp phase transitions it undergoes when the
average connectivity is varied. The problem translates in physics into a 
mean-field model with an ideal glass transition. The cavity method has been
efficient in giving insight into the problem, although the important challenge
of proving rigorously these results remains.  It would be interesting in this
respect to confirm these predictions by performing extensive Monte-Carlo
simulations of the phase diagram depicted in figure \ref{fig2}.

We also discussed the dynamical behavior of local search algorithms. We saw
that although the ``dynamical'' transition has a direct meaning as the point
where local Monte-Carlo algorithms get out of equilibrium, it is not directly
connected with the onset of hardness in the problem which is rather due to the
fact that clusters ``freeze'' as the connectivity increases.  So far, it has
not therefore been possible, despite initial hope, to have well-defined
HARD and EASY phases. The precise role of frozen variables and the part played by
the rigidity transition, or by the connectivity where all clusters becomes
frozen, will undoubtly be the subject of new research, both in
numerical and theoretical directions, that will help to clarify these issues. A
refined knowledge of the energy landscape would also be valuable.

Finally, we also discuss the major breakthrough in the algorithmic
strategy that emerged from the application of the cavity solution on a
single given graph. It is not clear at the present time what it the
best way to use this approach and how efficient it can be, nor if it
is possible to go arbitrarily close to the satisfiability/colorability
threshold for any values of $q$, and it is likely that these questions
will also trigger a lot of work in the future.

\ack We thank J. Kurchan, A. Montanari, F. Ricci-Tersenghi and G.  Semerjian
for the collaborations that led to a substantial part of our results. We also
greatly benefit from discussions with M.  M\'ezard and R. Zecchina. We thank
T.  J\"org and A. Hartmann for a critical lecture of the paper.

\end{document}